\documentclass[journal]{IEEEtran}
\usepackage{amsmath}
\usepackage{amsfonts}
\usepackage{amsthm}
\usepackage{cite}
\usepackage{soul,color}
\usepackage{graphicx}
\usepackage{float}
\usepackage{epstopdf}
\usepackage{amssymb}
\usepackage{stmaryrd}
\usepackage{multirow}

\newtheorem{notation}{\textit{Notations}}
\usepackage{enumitem}
\usepackage{algorithm}
\usepackage{algorithmic}
\usepackage{bibspacing}
\begin{document} 
\title{Deep Learning Based Proactive Optimization for Mobile LiFi Systems with Channel Aging}
\author{Mohamed Amine Arfaoui$^{*}$,
        Ali Ghrayeb, 
        Chadi Assi,
        and Marwa Qaraqe
        \vspace{-0.7cm}
\thanks{Mohamed Amine Arfaoui and Chadi Assi are with the Concordia Institute for Information Systems Engineering (CIISE), Concordia University, Montreal, QC H3G 1M8, Canada (e-mail: m$\_$arfaou@encs.concordia.ca;
assi@ciise.concordia.ca).\\
\indent Ali Ghrayeb is with the Electrical and Computer Engineering (ECE) Department, Texas A\&M University at Qatar, Doha, Qatar (e-mail: ali.ghrayeb@qatar.tamu.edu).\\
\indent Marwa Qaraqe is with the College of Science and Engineering, Hamad Bin Khalifa University, Qatar Foundation, Doha, Qatar (e-mail:
mqaraqe@hbku.edu.qa).}
\thanks{$^*$\textit{Corresponding author: M.A. Arfaoui, m$\_$arfaou@encs.concordia.ca}}
}
\maketitle 
\thispagestyle{plain}
\begin{abstract} This paper investigates the channel aging problem of mobile light-fidelity (LiFi) systems. In the LiFi physical layer, the majority of the optimization problems for mobile users are non-convex and require the use of dual decomposition or heuristics techniques. Such techniques are based on iterative algorithms, and often, cause a high processing delay at the physical layer. Hence, the obtained solutions are no longer optimal since the LiFi channels are evolving. In this paper, a proactive-optimization (PO) approach that can alleviate the LiFi channel aging problem is proposed. The core idea is to design a long-short-term-memory (LSTM) network that is capable of predicting posterior positions and orientations of mobile users, which can be then used to predict their channel coefficients. Consequently, the obtained channel coefficients can be exploited to derive near-optimal transmission-schemes prior to the intended service-time, which enables real-time service. Through various simulations, the performance of the designed LSTM model is evaluated in terms of prediction error and time, as well as its application in a practical LiFi optimization problem. 
\end{abstract} 
\section{Introduction}
\indent As the fifth generation (5G) cellular systems are currently under deployment, researchers from both academia and industry started shaping their vision on the upcoming sixth generation (6G) \cite{saad2019vision}. The main goals of 6G networks are to fill the gap of the original and unfulfilled promises of 5G and to keep up with the continuous emergence of the Internet of-Things (IoTs) networks \cite{zhang20196g}. Therefore, 6G networks must urgently provide high data rates, seamless connectivity, ubiquitous coverage, and ultra-low latency communications in order to reach the preset targets \cite{zhang20196g}. Due to this, researchers from both industry and academia are trying to explore new network architectures, new transmission techniques, and higher frequency bands, such as the millimeter wave (mmWave), the terahertz (THz), the infrared (IR), and the visible light (VL) bands, to meet these high requirements \cite{zhang20196g}. \\
\indent Light-fidelity (LiFi) is a bidirectional, high speed, and fully networked optical wireless communication (OWC) technology, that uses VL as the propagation medium in the downlink for the purposes of illumination and wireless communication. It also uses IR in the uplink so that the illumination constraint of a room remains unaffected, and also to avoid interference with the VL in the downlink \cite{haas2015lifi}. LiFi offers a number of important benefits that have made it favorable for 6G networks, such as the very large, unregulated bandwidth available in the VL and IR spectra \cite{david20186g}, the high energy efficiency \cite{tavakkolnia2018energy}, and the enhanced security as light does not penetrate through opaque objects \cite{arfaoui2020physical}. \\
\indent Unlike in conventional radio frequency wireless systems, the OWC channel is not isotropic, meaning that the positions and orientations of the optical transmitters and receivers affects the channel gain significantly \cite{arfaoui2020measurements}. Due to this, in the context of LiFi technology, the joint knowledge of UE position and orientation is a crucial factor for channel estimation and resource management tasks \cite{arfaoui2021invoking}. For this purpose, novel and accurate position and orientation estimation solutions were recently proposed in the literature, such as in \cite{yin2015indoor,zhou2019joint,arfaoui2021invoking} and references therein. However, these techniques are limited to the case of static UEs, and hence, they can not be applied to the case of mobile UEs as explained in the following. \\
\indent By definition, a LiFi UE is mobile if its position and/or orientation are changing over consecutive time slots, such as the case of a user that is browsing or watching a streaming video while walking \cite{soltani2018modeling}. On the other hand, the majority of the invoked optimization problems at the LiFi physical layer are non-convex, e.g., maximizing throughput by means of power control and optimal precoding for sum rate maximization, to name only a few \cite{ali20206g}. These problems may be solved using dual decomposition or heuristic techniques that require iterative algorithms, which induce a certain processing time at the LiFi physical layer \cite{ali20206g}. However, for the case of mobile LiFi UEs, such processing delay may exceed the maximum amount of time allocated to serve all the mobile UEs, and therefore, can not be tolerated. In fact, since users are mobile, they might change their instantaneous positions and/or orientation within the processing time. Consequently, their channel coefficients may evolve within the processing time and their previously estimated channel coefficients are outdated and no longer accurate. This makes the obtained solution no longer optimal after the processing time, which can lead to performance degradation. This problem is known as \textit{channel aging}, i.e., the LiFi channel coefficients are outdated after the optimization time, which is a very known problem in the wireless and mobile communication literature \cite{papazafeiropoulos2016impact,yuan2020machine}.\\
\indent In this paper, a proactive optimization (PO) approach is proposed to overcome the aforementioned \textit{channel aging} problem. Considering a certain channel-based optimization problem with respect to a certain transmission strategy and a certain performance metric, the proposed PO approach consists of proactively solving the considered problem and determining near-optimal schemes prior to the intended service time. Given a certain number of prior time slots $N$ and a certain number of posterior time slots $L$, the operation of the proposed PO approach can be summarized as follows. First, at each time slot $t>0$, the LiFi controller collects $N$ signal-to-noise ratio (SNR) values for each LiFi UE during the time interval $\left[t-N+1,t \right]$. Second, the collected SNR values of each LiFi UE are fed into a prediction unit that can predict its 3D position and orientation associated to the posterior time slot $t+L$, which are then exploited to predict its associated channel coefficients relative to time slot $t+L$. Afterwards, the predicted channel coefficients of the coexistent LiFi UEs are fed into the considered optimization problem in order to solve it within the time interval $\left[t,t+L \right]$. Finally, a near-optimal solution can be obtained prior to the intended service time slot $t+L$ without any processing delay, and therefore, the aforementioned \textit{channel aging} problem can be alleviated. \\ 
\indent One key component in the proposed PO approach is the prediction unit that can predict the posterior 3D position and orientation of each UE based on its prior received SNR values. For this task, deep learning (DL) techniques are invoked. Specifically a long-short-term-memory (LSTM) network is designed and trained offline in order to map the prior received SNR values with their associated posterior 3D positions and orientations. In the online phase, the obtained LSTM model is deployed at the APs. The APs receive signals from each UE, calculate its SNR values, and then the APs controller applies the trained LSTM model to predict the posterior 3D position and orientation of the UE. \\ 
\indent The performance of the designed LSTM model, in terms of prediction error, precision, and computational time, is compared with an optimized CNN model, which is the best ML-based approach in estimating the 3D position and orientation of LiFi UEs reported in the literature \cite{arfaoui2021invoking}. In addition, as an application of the proposed PO approach, a typical optimization problem is considered, which aims at maximizing the sum-rate of a multi-user multiple-input single-output (MISO) mobile LiFi system with respect to the weights of zero-forcing (ZF) precoding technique, while guaranteeing a certain quality of service (QoS) for each LiFi UE. The simulation results demonstrate that the proposed PO approach is able to provide a near-optimal and real-time service of the mobile LiFi UEs. \\
\indent The rest of the paper is organized as follows. The system model and problem statement are presented in Section II. Section III presents the proposed PO approach. Sections IV presents the prediction of the 3D position and orientation of LiFi UEs. Sections V and VI present the simulation results and the conclusion, respectively. 
\section{System Model and Problem Statement} 
\label{sec:system_model}
\subsection{System Setup}
\begin{figure}[t]
\centering     
\includegraphics[width=0.8\linewidth]{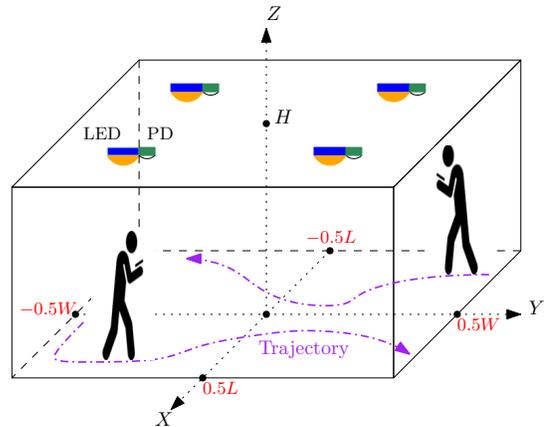}
\caption{A typical indoor LiFi system.}
\label{fig:IndEnv}
\end{figure}

We consider the indoor LiFi system shown in Fig.~\ref{fig:IndEnv}, which consists of $M$ APs installed at the ceiling. Each AP is equipped with one LED and one PD adjacent to each other, where the LED is used for illumination and downlink data transmission simultaneously and the PD is used for uplink data reception. Within this LiFi system, $K$ mobile LiFi users are communicating simultaneously (within the same time/frequency resource block) with the APs, where each user is equipped with a LiFi-enabled UE and is moving following a certain trajectory. Each UE is equipped with a single IR LED and a single PD that are used for data transmission and reception in the uplink and downlink phases, respectively. In addition, each UE has a random orientation over time, i.e., at each point of the trajectory of each user, the orientation of its associated UE is random. 
\subsection{Transmission Model}
\indent In this system model, we focus on the downlink transmission, where the intensity-modulation direct-detection (IM/DD) is considered. For all $k \in \llbracket 1, K \rrbracket$, the downlink received signal at the $k$th UE at each time slot $t>0$ is expressed as \cite{mohammad2019optical},
\begin{equation}
\label{eq:downlink_received_signal}
y_k(t)= \mathbf{h}_{k}^T(t)\mathbf{s}(t)+n_k(t),
\end{equation}
 where $\textbf{h}_{k}(t)$ represents the instantaneous downlink $M \times1$ channel gain vector between the APs and the $k$th UE, $\mathbf{s}(t)$ represents the $M \times 1$ information-bearing signal broadcast by the APs at time slot $t$, which contains a mixture of the data of the $K$ users, and $n_k(t)$ is the instantaneous additive white Gaussian noise (AWGN) experienced at the PD of the $k$th UE, which is $\mathcal{N}\left(0, N_0 B \right)$ distributed, where $N_0$ is the noise power spectral density and $B$ is the bandwidth of the system. \\ 
\indent The performance of the considered LiFi system depends heavily on the instantaneous $K \times M$ channel matrix $\mathbf{H}(t) = \left[\mathbf{h}_{1}(t),\mathbf{h}_{2}(t),...,\mathbf{h}_{K}(t) \right]^T$, which in turn depends on the positions of the APs as well as the instantaneous 3D position and orientation of each UE \cite{arfaoui2021invoking}. Basically, for $k \in \llbracket 1, K \rrbracket$, the instantaneous 3D position of the $k$th user is characterized by its instantaneous coordinates $\left(x_k(t),y_k(t),z_k(t)\right)$ in the Cartesian coordinate system $\left(X,Y,Z \right)$ shown in Fig.~\ref{fig:IndEnv}, whereas its instantaneous 3D orientation is fully characterized through the instantaneous elemental rotation angles with respect to the three axis of the UE, which are the yaw angle $\alpha_k(t) \in [0^\circ,360^\circ)$, the pitch angle $\beta_k(t) \in [-180^\circ,180^\circ)$, and the roll angle $\gamma_k(t) \in [-90^\circ,90^\circ)$ \cite{arfaoui2021invoking, soltani2018modeling}.
\subsection{Problem Statement}
\label{sec:problem-statement}
\indent Let us consider an optimization problem $\mathcal{P}\left[\mathbf{H}(t)\right]$ that aims at enhancing the performance of the LiFi system at hand at each time slot $t$, with respect to a certain performance metric, such as sum-rate maximization, minimum rate maximization, latency minimization, etc, \cite{pham2020energy,zhao2020multiuser,ma2015coordinated,arfaoui2019secrecy,arfaoui2018artificial,arfaoui2018secrecy}. This class of problems may be solved using dual decomposition techniques or heuristic approaches that require iterative algorithms. In this case, let $\Delta t$ denote the processing time required to solve the optimization problem $\mathcal{P}\left[\mathbf{H}(t)\right]$ at each time slot $t$, which increases as the number of mobile users and/or APs increases. Within the time interval $\Delta t$, the channel matrix of $\mathbf{H}$ evolves from $\mathbf{H}(t)$ to $\mathbf{H}(t+\Delta t)$, since the users are mobile and may have changed their instantaneous positions and/or orientations during the time interval $\Delta t$. In this case, the obtained solution from solving problem $\mathcal{P}\left[\mathbf{H}(t)\right]$ will be applied at time $t+\Delta t$ and, therefore, it is no longer optimal. In other words, the obtained solution, which is optimal for $\mathbf{H}(t)$, is not optimal for $\mathbf{H}(t+\Delta)$ and can imply a performance loss to the system. This problem is known as \textit{channel aging}, i.e., the channel matrix $\mathbf{H}(t)$ is outdated at time slot $t+\Delta t$, and this is actually a very known problem in the wireless and mobile communication literature, especially for massive MIMO systems \cite{papazafeiropoulos2016impact,yuan2020machine}. Based on the above discussion, we propose in the following section a PO approach to overcome the aforementioned \textit{channel aging} problem.
\section{Proposed PO Approach} 
\label{sec:proposed_approach}
In this section, the proposed PO technique is presented. The basics of the proposed approach are discussed first, and then, its detailed implementation is investigated.
\subsection{Core Idea}
Let $L\in \mathbb{N} \backslash \{0\}$ denote a posterior time slot index and let us consider the optimization problem $\mathcal{P}\left[\mathbf{H}(t+L)\right]$ that needs to be solved at each time slot $t+L$. In order to overcome the \textit{channel aging} problem discussed in Section \ref{sec:problem-statement}, an alternative approach is to solve problem $\mathcal{P}\left[\mathbf{H}(t+L)\right]$ prior to the occurring of the target time slot $t+L$. In this context, the proposed PO approach consists first of predicting the channel matrix $\mathbf{H}\left(t+L\right)$ at time slot $t$, which is denoted by $\widehat{\mathbf{H}}\left(t+L\right)$. Then, the optimization problem $\mathcal{P}\left[\widehat{\mathbf{H}}(t+L)\right]$ is solved within the time interval $\left[t, t + L \right]$. In this case, a sub-optimal solution is obtained and can be employed in serving the cellular users at time slot $t+L$, without any processing delay at time slot $t+L$, which will will suppress indeed the raised \textit{channel aging} problem. \\
\begin{figure}[t]
        \centering
        \includegraphics[width=1\linewidth]{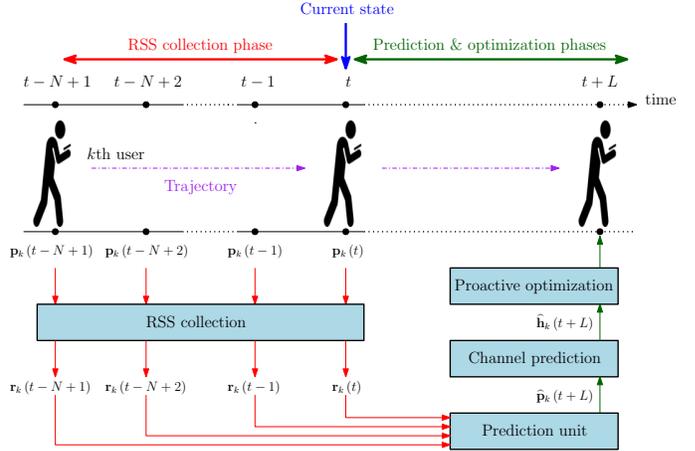}
        \caption{Procedure of the proposed PO technique.}
        \label{fig:methodology}
\end{figure}
\indent Fig.~\ref{fig:methodology} presents the procedure of the proposed PO approach. As shown in this figure, it consists mainly of three consecutive phases, which are: 1) SNR collection phase, 2) Prediction phase, and 3) Optimization phase. For the ease of reading, we start with the following notations.
\begin{notation}
For all $k \in \llbracket 1, K \rrbracket$, we denote by $\mathbf{p}_k(t) \triangleq \left[x_k(t),y_k(t),z_k(t),\alpha_k(t),\beta_k(t),\gamma_k(t) \right]$ the $1\times 6$ vector that contains the exact instantaneous 3D position orientation and of the $k$th UE at time slot $t$. Accordingly, we denote by $\widehat{\textbf{p}}_k(t)$ the predicted 3D position orientation of the $k$th UE at time slot $t$, respectively.
\end{notation}
Now, the details of each phase of the proposed PO approach are presented in the following subsections. 
\subsection{SNR Collection Phase}
\label{sec:SNR}
\indent In this phase, $N \in \mathbb{N} \backslash \{0\}$ uplink SNR values for each UE are collected over the time slots $\llbracket t-N+1, t \rrbracket$. Specifically, for all $k \in \llbracket 1, K \rrbracket$, the $k$th UE broadcasts through its IR-LED a scalar reference signal $s_k = I_{\rm DC}$, where $I_{\rm DC}$ denotes the direct-current (DC) intensity bias of the IR-LED. Moreover, in order to cancel the uplink inter-user interference in this phase, the total bandwidth $B$ is equally divided over all $K$ UEs for the uplink transmissions. Hence, for all $k \in \llbracket 1, K \rrbracket$ and $i \in \llbracket 1, M \rrbracket$, the received signal of the $k$th UE at the $i$th AP is given by $z_{k,i} = g_{k,i}(j)  s_k + \omega_{i}$, where $g_{k,i}(j)$ is the total uplink channel gain between the $k$th UE and the $i$th AP at time slot $j$, and $\omega_{i}$ is an AWGN experienced at the $i$th AP that is $\mathcal{N}\left(0, \sigma_{\rm u}^2 \right)$ distributed, such that $\sigma_{\rm u}^2 = \frac{N_0 B}{K}$. Consequently, for all $k \in \llbracket 1, K \rrbracket$ and $i \in \llbracket 1, M \rrbracket$, the received SNR of the $k$th UE at the $i$th AP at each time slot $j \in \llbracket t-N+1, t \rrbracket$ is expressed as
\begin{equation}
    \label{eq:SNR}
    r_{k,i}(j) = \frac{g_{k,i}^2(j)I_{\rm DC}^2}{\sigma_u^2}.
\end{equation}
Based on this, for all $k \in \llbracket 1, K \rrbracket$, the $1 \times M$ received SNR vector of the $k$th UE at the APs at each time slot $j \in \llbracket t-N+1, t \rrbracket$ is expressed as 
\begin{equation}
\mathbf{r}_k(j) \triangleq \left[r_{k,1}(j), r_{k,2}(j),...,r_{k,M}(j) \right],
\end{equation}
which depends mainly on the instantaneous 3D position and orientation vector of the $k$th UE at time slot $j$, that is denoted by $\mathbf{p}_k(j)$. Consequently, as shown in Fig.~\ref{fig:methodology}, within the time interval $\llbracket t-N+1, t \rrbracket$ and for all $k \in \llbracket 1, K \rrbracket$, the APs collect a group of $N$ SNR vectors for the $k$th UE that is expressed as
\begin{equation}
    \mathbf{r}_{k, \rm total}(t) = \left[\mathbf{r}_k(t-N+1),\mathbf{r}_k(t-N+2),...,\mathbf{r}_k(t) \right].
\end{equation}
\subsection{Prediction Phase} 
\label{sec:prediction_phase}
In this phase, and as shown in Fig.~\ref{fig:methodology}, the goal is to predict, at each time slot $t$, the vector $\mathbf{p}_k(t+L)$ that contains the posterior 3D position and orientation for each UE relative to time slot $t+L$, based on its associated vector of prior SNR values $\mathbf{r}_{k, \rm total}(t)$ collected between time slots $t-N+1$ and $t$. As was mentioned in the SNR collection phase in subsection \ref{sec:SNR}, for all $k \in \llbracket 1, K \rrbracket$, the vector of SNR values $\mathbf{r}_{k, \rm total}(t)$ collected between time slots $t-N+1$ and $t$ is a function of the instantaneous 3D position orientation vector $\mathbf{p}_k(j)$ of the $k$th UE at each time slot $j \in \llbracket t-N+1, t \rrbracket$. In other words, at each time slot $t$ and for all $k \in \llbracket 1, K \rrbracket$, the vector of total SNR values $\mathbf{r}_{k, \rm total}(t)$ collected between the time slots $t-N+1$ and $t$ contains information about the previous realizations of the 3D position and orientation of the $k$th UE within the time interval $\llbracket t-N+1, t \rrbracket$. Therefore, there exists a deterministic vector-valued function $\mathbf{F} \left(\cdot \right)$ such that 
\begin{equation}
        \label{eq:estimation}
		\mathbf{p}_k(j) = \mathbf{F} \left(\mathbf{r}_k(j)\right). \\
\end{equation}
\indent On the other hand, for all $k \in \llbracket 1, K \rrbracket$, since the $k$th UE is mobile, then its instantaneous 3D position and orientation vector $\mathbf{p}_k$ is a multivariate random process (RP). In this case, for all $k \in \llbracket 1, K \rrbracket$ and for each time slot $t$, there exists a probabilistic vector-valued function $\mathbf{G} \left( \cdot \right)$, such that 
\begin{equation}
    \label{eq:random_process}
    \begin{split}
    &[\mathbf{p}_k(t+L), \mathbf{p}_k(t+L-1), \cdots, \mathbf{p}_k(t)] \\
    &\qquad \qquad = \mathbf{G} \left(\mathbf{p}_k(t), \mathbf{p}_k(t-1), \cdots, \mathbf{p}_k(t-N+1)\right).
    \end{split}
\end{equation}
Consequently, based on \eqref{eq:estimation} and \eqref{eq:random_process}, there exists a probabilistic vector-valued function $\mathbf{J} \left( \cdot \right)$, such that, for all $k \in \llbracket 1, K \rrbracket$ and for each time slot $t$,
\begin{equation}
    \label{eq:mapping}
    \begin{split}
    &[\mathbf{p}_k(t+L), \mathbf{p}_k(t+L-1), \cdots, \mathbf{p}_k(t)] \\
    &\qquad \qquad = \mathbf{J} \left(\mathbf{r}_k(t), \mathbf{r}_k(t-1), \cdots, \mathbf{r}_k(t-N+1)\right).
    \end{split}
\end{equation}
Based on \eqref{eq:mapping}, the objective here is to determine the probabilistic vector-valued function $\mathbf{J} \left( \cdot \right)$. However, obtaining its exact characterization is not straightforward. In fact, for all $k \in \llbracket 1, K\rrbracket$ and for every time slot $j \in \llbracket t-N+1, t\rrbracket$, the vector of SNR values $\mathbf{r}_{k}(j)$ includes the contributions of both the line-of-sight (LOS) and the non-line-of-sight (NLOS) components of the wireless links between the $k$th UE and the $M$ APs. Although the contributions of the NLOS components will improve the prediction performance, as it was demonstrated in the joint 3D position and orientation estimation problem invoked \cite{arfaoui2021invoking}, their inclusion in the prediction process is not straightforward from an optimization point of view, like the case of maximizing an observed predictive likelihood function. This is basically due to their complex expressions as explained in \cite{arfaoui2021invoking}. To overcome this issue, an approximate parametric vector-valued function $\widehat{\mathbf{J}}\left( \mathcal{W},\cdot\right)$ should be constructed, where $\mathcal{W}$ is the associated set of parameters. In this case, for all $k \in \llbracket 1, K \rrbracket$ and at each time slot $t$, the predicted values of the posterior 3D position and orientation of the $k$th UE relative to time slot $t+L$ can be obtained as 
\begin{equation}
    \label{eq:mapping_approx}
    \begin{split}
    &[\widehat{\mathbf{p}}_k(t+L), \widehat{\mathbf{p}}_k(t+L-1), \cdots, \widehat{\mathbf{p}}_k(t)] \\
    &\qquad = \widehat{\mathbf{J}} \left(\mathcal{W}, \mathbf{r}_k(t), \mathbf{r}_k(t-1), \cdots, \mathbf{r}_k(t-N+1)\right)
    \end{split}
\end{equation}
The details on how the approximate parametric vector-valued function $\widehat{\mathbf{J}}$ and the optimal set of parameters $\mathcal{W}^*$ are obtained will be presented in section \ref{sec:prediction_section}. \\ 
\indent Based on the above, for all $k \in \llbracket 1, K \rrbracket$, the predicted 3D position and orientation vector $\widehat{\mathbf{p}}_k(t+L)$ of the $k$th UE associated to time slot $t+L$ is obtained and its associated predicted $M \times 1$ channel gain vector $\widehat{\textbf{h}}_k(t+L)$ can be calculated at time slot $t$. Consequently, the predicted $K \times M$ channel matrix $\widehat{\mathbf{H}}(t+L) = \left[\widehat{\textbf{h}}_1(t+L),\widehat{\textbf{h}}_2(t+L),...,\widehat{\textbf{h}}_K(t+L) \right]^T$ can be obtained at time slot $t$.
\subsection{Optimization Phase}
In this phase, once the predicted $K \times M$ channel matrix $\widehat{\mathbf{H}}(t+L)$ is obtained, the optimization problem $\mathcal{P}\left[\widehat{\mathbf{H}}(t+L)\right]$ is solved within the time interval $\left[t, t + L \right]$. In this case, a sub-optimal solution is obtained and can be employed in serving the cellular users at time slot $t+L$, without any processing delay at time slot $t+L$, which overcomes the problem of \textit{channel aging} of the considered LiFi system that was raised in subsection \ref{sec:problem-statement}.
\section{Joint Prediction of Indoor LiFi user Position and Orientation}
\label{sec:prediction_section}
\subsection{Methodology}
\indent In this section, and as discussed in subsection \ref{sec:prediction_phase}, our objective is to determine an approximate parametric vector-valued function $\widehat{\mathbf{J}}$ along with its optimal set of parameters $\mathcal{W}^*$ that can predict the posterior 3D position and orientation of the LiFi UEs with a good accuracy. To reach this goal, DL techniques, through the use of artificial neural networks (ANNs), are employed for this task. DL is a particular machine learning technique that implements the learning process elaborating the data through ANNs. The use of ANNs is a key factor that makes DL outperform other machine learning schemes, especially when a large amount of data is available \cite{zappone2019wireless}. This has made DL the leading ML technique in many scientific fields such as image classification, text recognition, speech recognition, audio and language processing and robotics \cite{zappone2019wireless}. The potential application of DL to physical layer communications has also been increasingly recognized because of the new features for future communications, such as complex scenarios with unknown channel models, high speed and accurate processing requirements, which present big challenges to 6G wireless networks \cite{wang2017deep}. Motivated by this, DL has been applied to wireless communications, such as physical layer communications \cite{wang2017deep,qin2019deep}, resource allocation \cite{ye2019deep,tang2018novel}, and intelligent traffic control \cite{tang2017removing}. Motivated by the above discussion, DL techniques are auspicious candidates for the prediction of posterior 3D position and orientation of the LiFi UEs, which is the focus of this section. \\
\indent The proposed prediction technique consists of two phases: 1) an offline learning (offline phase) and 2) an online deployment (online phase). In the offline learning, a dataset of $Q$ random sequences, each of length $N+L_{\rm max}$, is generated, where $L_{\max}$ denotes the maximum posterior time slot index. Specifically, for all $q \in \llbracket 1, Q \rrbracket$, a random trajectory of length $N+L_{\rm max}$ steps is generated using a predefined experimental-based indoor mobility model, within which a random 3D experimental-based orientation is generated at each step.  For all $q \in \llbracket 1, Q \rrbracket$, the process of generating the $q$th data point of the dataset is as follows:
\begin{enumerate}
\item A sequence $\left[\mathbf{p}_1^q,\mathbf{p}_2^q,...,\mathbf{p}_{N+L_{\max}}^q \right]$ of $N+L_{\max}$ 3D position and orientation vectors is randomly generated using the experimental-based indoor orientation-based random waypoint (ORWP) mobility model adopted in \cite{mohammad2019optical}, where for all $n \in \llbracket 1, N+L_{\max} \rrbracket$, $\mathbf{p}_n^q$ is the 3D position and orientation vector of the $n$th time step of the $q$th sequence. 
\item For all $n \in \llbracket 1, N \rrbracket$, the $1 \times M$ SNR vector $\mathbf{r}_{n}^q$ is calculated based on the 3D position and orientation vector $\mathbf{p}^q(n)$. 
\item The features vector of the $q$th data point is the $1 \times N$ sequence of SNR vectors $\left[\mathbf{r}_{1}^q,\mathbf{r}_{2}^q,...,\mathbf{r}_{N}^q \right]$. 
\item The labels vector of the $q$th data point is the $1 \times L$ sequence of posterior 3D position and orientation vector $\left[\mathbf{p}_{N+1}^q,\mathbf{p}_{N+2}^q,...,\mathbf{p}_{N+L_{\max}}^q \right]$. 
\end{enumerate}

\indent The generated features and labels are then stored into a single dataset. Afterwards, based on the obtained dataset, ANN models are designed and trained in order to construct the optimal model that provides the best approximate parametric vector-valued function $\widehat{\mathbf{J}}$ that can map between the received SNR vectors and the posterior 3D position and the orientation vectors are designed. In the following, we will present first the adopted ANN model and then we will discuss the deployment of the trained ANN model in the online phase.
\subsection{Prediction Model} 
\indent The problem at hand is the prediction of future 3D positions and orientations of a LiFi UE based on its prior received SNR values. Since both the position and the orientation of the UE can be modeled as RPs, as discussed in the previous paragraph, the problem can be formulated as a sequence-to-sequence (Seq2Seq) prediction problem \cite{shen2020sequence,hua2019deep}. Seq2Seq prediction is basically a process of extracting useful information from historical records and then determining future values. Unlike regression predictive modeling, Seq2Seq mapping adds the complexity of sequences dependencies among the input variables \cite{shen2020sequence,hua2019deep}. A powerful type of neural network designed to handle sequences dependencies is called recurrent neural networks (RNNs). The LSTM network is a category of RNN that is trained using Backpropagation through time and that is known of their capability in alleviating the vanishing gradient problem \cite{sundermeyer2012lstm}. As such, LSTM networks can be used to create large RNNs that can be used to address difficult Seq2Seq prediction problems. Motivated by this discussion, the approximate parametric vector-valued function $\widehat{\mathbf{J}}$ is an LSTM network that is trained in the offline phase using the generated dataset. Hence, its optimal set of parameters $\mathcal{W}^*$ that provides the best prediction performance can be obtained. 
\subsection{Online Phase}
In the online phase, once the LSTM model is optimized, the obtained parametric vector-valued function $\widehat{\mathbf{J}} \left(\mathcal{W}^*,\cdot \right)$ is deployed at the APs. Consequently, for all $k \in \llbracket 1, K \rrbracket$ and at each time slot $t$, the APs track if there is any change in the received SNR values $\left[\mathbf{r}_{k}(t-N+1),\mathbf{r}_{k}(t-N+2),...,\mathbf{r}_{k}(t) \right]$ from the $k$th UE within the interval of time $\llbracket t-N+1,t\rrbracket$. If this is the case, the received SNR values are fed into the obtained parametric vector-valued function $\widehat{\mathbf{J}} \left(\mathcal{W}^*,\cdot \right)$ in order to output the sequence of predicted posterior 3D position and orientation vectors of the $k$th UE $\left[\widehat{\mathbf{p}}_{k}(t),\widehat{\mathbf{p}}_{k}(t+1),...,\widehat{\mathbf{p}}_{k}(t+L) \right]$, where $L \in \llbracket 1, L_{\max} \rrbracket$ is the target posterior time slot index. Consequently, the obtained sequences of predicted posterior 3D positions and orientations vectors are injected into the channel matrix expression in order to predict the channel matrix of all UEs associated to time slot $t+L$, i.e., $\widetilde{\textbf{H}}(t+L)$. Finally, assuming a certain transmission strategy and a predefined performance metric, the associated optimization problem $\mathcal{P}\left[\widehat{\mathbf{H}}(t+L)\right]$ is solved within the time interval $\left[t, t + L \right]$. In this case, a sub-optimal solution is obtained and can be employed in serving the cellular users at time slot $t+L$ directly, without any processing delay at time slot $t+L$.
\section{Simulation Results}
In this section, our objective is to evaluate the performance of the proposed PO approach through extensive simulations.
\subsection{Simulations Parameters}
\begin{table}[t]
\caption{LSTM Specification}
\centering
\renewcommand{\arraystretch}{1} 
\setlength{\tabcolsep}{0.45cm} 
\begin{tabular}{| c | c |}
  \hline
  Prior time slot index & $N = 8$ \\ 
  \hline 
  Maximum posterior time slot index & $L_{\max} = 4$ \\
  \hline
  Dataset size & $Q = 10^6$ \\ 
  \hline 
  (Train, test) partition & $(0.9,0.1)\times Q$ \\
  \hline 
  Number of LSTM units & $100$ \\ 
  \hline 
  LSTM activation function & Hyperbolic tangent (tanh) \\ 
  \hline
  LSTM recurrent activation function & Sigmoid \\ 
  \hline
\end{tabular} 
\label{table:T1}
\end{table}
\indent In this section, we consider a typical indoor environment with dimensions $L \times W \times H$ = $5\times 5 \times 3$ m$^3$. The LiFi system is equipped with $M = 16$ APs that are arranged on the vertices of a square lattice over the ceiling of the room, where each AP is oriented vertically downward. A LiFi UE, that is equipped with one IR-LED and one PD adjacent to each other, is moving within the room and its UE may have a random orientation along the random trajectory. The UE is a typical smartphone with dimensions $14 \times 7 \times 1$ cm$^3$, where the IR-LED and the PD are placed at the screen of the smartphone, exactly at $6$ cm above the center. The simulation parameters along with the LOS and NLOS channel models are adopted from \cite{arfaoui2021invoking}. The specification of the proposed LSTM model is summarized in Table \ref{table:T1}. The central processing unit (CPU) of the machine on which all the simulations were performed was an Intel Core i5 from the second generation that has a dual-core, a basic frequency of 2.40 GHz and a maximum turbo frequency of 3.40 GHz.
\subsection{Learning, Prediction, and Complexity Performance Evaluation}
\begin{figure}[t]
       \centering
        \includegraphics[width=0.9\linewidth]{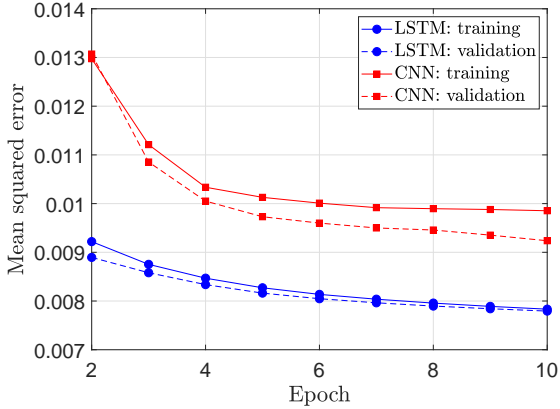}
        \caption{Training and validation losses versus the epoch index.}
        \label{fig:trainning_validation_loss}
\end{figure}

\indent We adopt the convolutional neural network (CNN), which is so far the best 3D position and orientation estimation technique proposed in the literature \cite{arfaoui2021invoking}, as a baseline to assess the performance of our proposed LSTM model. Fig.~\ref{fig:trainning_validation_loss} presents the training and validation losses of the designed LSTM and CNN networks, measured in terms of the mean-squared-error (MSE), versus the training epoch index. The training and validation losses decrease as the epoch index increases, which demonstrates that the designed LSTM and CNN networks are both converging, not overfitting, and can generalize well over unseen data points in the online phase. \\
\begin{table}[t]
\caption{Average positioning error.}
\centering
\renewcommand{\arraystretch}{1} 
\setlength{\tabcolsep}{0.9cm} 
\begin{tabular}{| c | c | c |}
  \cline{2-3} 
   \multicolumn{1}{c|}{} & LSTM & CNN \\
  \hline 
   $L=1$ & $0.1789$ m & $0.4742$ m \\
   \hline
   $L=2$ & $0.2136$ m & $0.5192$ m \\
   \hline 
   $L=3$ & $0.2565$ m & $0.5750$ m \\
   \hline
   $L=4$ & $0.3046$ m & $0.6385$ m \\
   \hline
\end{tabular} 
\label{table:T2}
\end{table}
\begin{figure}[t]
\centering
\includegraphics[width=0.9\linewidth]{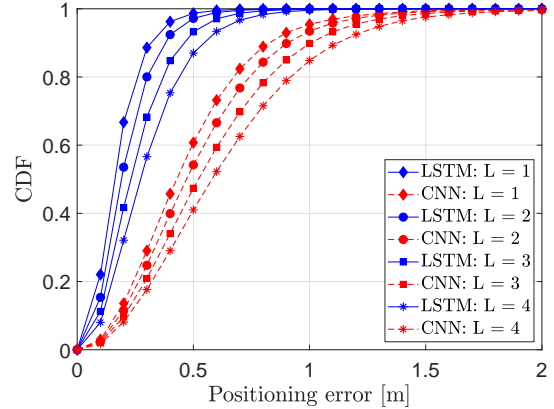}
\caption{CDF of the positioning error.}
\label{fig:position_cdf}
\end{figure}
\indent Table \ref{table:T2} presents the average prediction error of the position for the proposed LSTM model and the considered CNN model for different values of the time slot index $L$. In addition, Fig.~\ref{fig:position_cdf} presents the empirical cumulative distribution function (CDF) of the instantaneous positioning error resulting from the designed LSTM and CNN models for different values of the posterior time slot index $L$. Table \ref{table:T2} and  Fig.~\ref{fig:position_cdf} show that the LSTM model provides higher positioning accuracy than the adopted CNN baseline. Moreover, the results of Table \ref{table:T2} and  Fig.~\ref{fig:position_cdf} show that the instantaneous position error increases as the posterior time slot index increases. This result is expected since, as the posterior time slot index $L$ increases, more uncertainty about the future positions will be considered. The same results also applies to the prediction performance of the orientation angles yaw $\alpha$, pitch $\beta$ and roll $\gamma$, which were omitted here for space limitation.\\
\indent The required time to generate a dataset of size $Q = 10^6$ data points is approximately $6$ hours. In the other hand, the time required to train the designed LSTM and CNN models is $14$ mins and $1.5$ hours, respectively. Although the time required for the dataset generation and the models training is large, this high computational time is not an issue, since the dataset generation and the models training is performed in the offline phase and only once prior to the deployment of the APs. Considering the online complexity, the computational time of the designed LSTM and CNN models in the online phase is extremely low. In fact, over the whole test set, which has a size of $0.1 \times Q = 10^5$, the total prediction time in the online phase is $3$ seconds and $6.56$ seconds for the LSTM and CNN models respectively. Therefore, the average prediction time per trial is $\frac{3}{10^5} = 0.03$ millisecond and $\frac{6.56}{10^5} = 0.07$ millisecond for the LSTM and CNN models respectively, i.e., real-time prediction. These results, along with the prediction accuracy results, demonstrate the superiority of the designed LSTM model over the CNN model, which makes it an auspicious solution for accurate and real-time prediction. 
\subsection{Application: Proactive Sum Rate Maximization for Multi-User MISO Mobile LiFi System with QoS Constraints.}
\indent As an application of the proposed PO approach, we consider in this section the proactive sum-rate maximisation of a multi-user MISO mobile LiFi system. A MISO LiFi system is serving multiple coexisting mobile users simultaneously, and in order to cancel the inter-user interference, the ZF precoding is applied. The objective is to maximize the instantaneous sum-rate of the system with respect to the ZF precoder, while respecting the peak-power constraint at the LiFi APs and guaranteeing a certain QoS for each user that is specified in terms of a minimum rate threshold $R_{\rm th}$. This problem is a channel-based problem, since the ZF precoder is a function of the channel matrix of the system. In addition, this problem was considered in the literature and was shown to be a non-convex one \cite{al2018multi,shen2016rate,pham2017multi,pham2018artificial,zhao2020multiuser}. Therefore, it suffers from the channel aging problem and the proposed PO approach must be applied.\\   
\indent The results presented within this section are obtained from independent Monte-Carlo trials over the whole test set. For a given prior time slot index $L$, and at each time slot $t$, four different cases are considered, which are
\begin{enumerate}
    \item \textit{Baseline 1}: The exact instantaneous channel matrix $\mathbf{H}\left(t+L\right)$ is perfectly estimated at time slot $t+L$ and then the corresponding optimal ZF precoder is obtained at the same time slot without considering any processing delay.
    \item \textit{Proposed PO}: The predicted instantaneous channel matrix $\widehat{\mathbf{H}}\left(t+L\right)$ is obtained from the optimized LSTM model at time slot $t$ and then the corresponding optimal ZF precoder is obtained within the time interval $\llbracket t, t+L \rrbracket$.
    \item \textit{Baseline 2}: The predicted instantaneous channel matrix $\widehat{\mathbf{H}}\left(t+L\right)$ is obtained from the optimized CNN model at time slot $t$ and then the corresponding optimal ZF precoder is obtained within the time interval $\llbracket t, t+L \rrbracket$.
    \item \textit{Baseline 3}: The exact instantaneous channel matrix $\mathbf{H}\left(t\right)$ is perfectly estimated at time slot $t$. Then the corresponding optimal precoder is obtained within the time interval $\llbracket t, t+L \rrbracket$ and then applied at time slot $t+L$.
\end{enumerate}
\indent \textit{Baseline 1} is considered in this section as an upper bound to assess the performance of \textit{Proposed PO} approach, since at each time slot $t+L$, the idealistic scenario is to obtain the corresponding optimal ZF precoder at the same time slot without any processing delay. Moreover, \textit{Baseline 3} is considered here to demonstrate the performance degradation caused by the \textit{channel aging} problem invoked in this paper, since the ZF precoder associated to the instantaneous channel matrix $\mathbf{H}\left(t\right)$ is obtained within the time interval $\llbracket t, t+L \rrbracket$ and then applied to the instantaneous channel matrix $\mathbf{H}\left(t+L\right)$. On the other hand, for each case discussed above, the associated ZF precoder is obtained using two approaches, which are 
\begin{itemize}
    \item The optimal approach that consists of running an off-the-shelf optimization solver. This technique is adopted from \cite{arfaoui2022comp} and references therein.
    \item The convex-concave approach (CCP). This technique is adopted from \cite{arfaoui2019secrecy} and references therein.
\end{itemize}
\begin{figure}[t]
        \centering
        \includegraphics[width=0.9\linewidth]{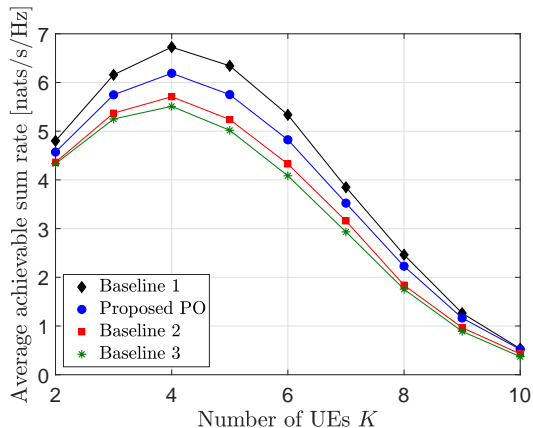}
        \caption{Sum-rate versus the number of LiFi UEs $K$, where the posterior time slot index $L=2$.}
        \label{fig:sum_rate_K}
\end{figure}

\begin{figure}
        \centering
        \includegraphics[width=0.9\linewidth]{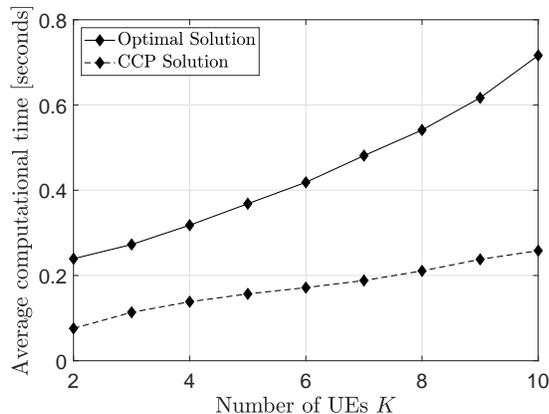}
        \caption{Average computational time versus the number of LiFi UEs $K$, where the required rate threshold for each LiFi UE $R_{\rm th} = 1$ nats/s/Hz.}
        \label{fig:time}
\end{figure}

\indent In line with the above, Fig.~\ref{fig:sum_rate_K} presents the average sum rate of the considered LiFi system versus the number of LiFi UEs $K$ for the different cases presented above, where the posterior time slot index $L=2$ and the target rate threshold (QoS) per UE is $R_{\rm th} = 1$ nats/s/Hz. Fig.~\ref{fig:sum_rate_K} demonstrates that the CCP approaches provides a near-optimal solution for the problem in hand, where the gap between average sum rate obtained from using the optimal solution and the CCP solution is less than $0.01\%$ for all the considered scenarios. On the other hand, by comparing \textit{Baseline 3} to \textit{Baseline 1}, Fig.~\ref{fig:sum_rate_K} highlights the performance degradation that is caused by the channel aging problem, where the peak gap between the average sum rates associated to \textit{Baseline 1} and \textit{Baseline 3} is higher than $20\%$ for the considered range of number of UEs. However, the \textit{proposed PO} approach outperforms \textit{Baseline 3} and brings a performance enhancement to the system, where the peak gap between the average sum rates associated to \textit{Baseline 1} and the \textit{proposed PO} is less than $7\%$. On the other hand, Fig.~\ref{fig:sum_rate_K} shows that the LSTM model outperforms the CNN model, since the peak gap between the average sum rates associated to \textit{Baseline 1} and \textit{Baseline 2} is higher than $10\%$. \\
\begin{figure}[t]
        \centering
        \includegraphics[width=0.9\linewidth]{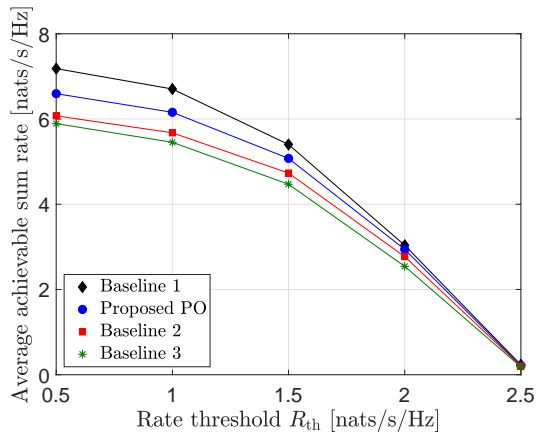}
        \caption{Sum-rate versus the required rate threshold for each LiFi UE $R_{\rm th}$, where the number of UEs is $K = 4$ and the posterior time slot index $L=2$.}
        \label{fig:sum_rate_rateth}
\end{figure}
\indent Fig.~\ref{fig:sum_rate_rateth} presents the average sum rate of the considered LiFi system versus the target rate threshold per UE $R_{\rm th}$, when the number of UEs is $K = 4$ and the posterior time slot index $L=2$. Basically, this figure shows the same conclusions derived from Fig.~\ref{fig:sum_rate_K}, which demonstrates the potential of the LSTM model in particular and the proposed PO approach in general. In addition, this figure shows that the sum rate of the system decreases as the rate threshold $R_{\rm th}$. This is basically expected, since when the rate threshold $R_{\rm th}$ increases, the number of users that are admitted to the networks decreases, and this is due to the fact that their channel realizations can not satisfy their target QoS. \\
\indent Fig.~\ref{fig:time} presents the average computational time of the optimal optimization approach and the CCP approach for obtaining the best ZF precoder for the considered LiFi system versus the number of LiFi UEs. This figure demonstrates that the CCP approach has a lower computational time than the one of the optimal optimization approach (at least $50\%$ less). In addition, recall that the CCP approach provides a performance gap that is less than $0.01\%$ from the optimal approach. This demonstrates the potential of the CCP approach in providing an optimal solution with a lower computational time for the considered problem. Moreover, assuming that the duration of one time slot is $0.5$ seconds, the solution of the CCP approach can be obtained within one time slot. Therefore, the proposed PO approach can be applied even for a posterior time slot index $L = 1$, i.e., at each time slot $t$, a near-optimal solution can be provided using the LSTM model and the CCP approach in the time interval $\llbracket t, t+1 \rrbracket$. Therefore, for each time slot $t$, a near optimal ZF precoder can be applied directly at time slot $t+1$ without any processing delay.
\section{Conclusion} 
With the goal of overcoming the LiFi channel aging problem, a PO approach for is proposed in this paper. The core of the proposed technique is an LSTM network that is capable of predicting posterior positions and orientations of mobile users, which are then used to predict the channel coefficients of mobile wireless links. Finally, the obtained predicted channel coefficients are exploited for deriving near-optimal power allocation schemes prior to the intended service time, which enables near-optimal and real-time service for mobile LiFi users. Through various simulations, the performance of the proposed PO approach is investigated and the obtained results demonstrated the potential of the proposed PO approach in alleviating the LiFi channel aging problem.
\bibliographystyle{IEEEtran}
\bibliography{main.bib} 
\end{document}